\documentclass[]{spie}  

\usepackage{amsmath,amsfonts,amssymb}
\usepackage{graphicx}
\usepackage{caption}
\usepackage{subcaption}
\usepackage[colorlinks=true, allcolors=blue]{hyperref}
\usepackage{color}

\title{Development of Calibration Strategies for the Simons Observatory} 

\author[a]{Sean A. Bryan}
\author[b]{Sara M. Simon}
\author[c]{Martina Gerbino}
\author[d]{Grant Teply}
\author[e]{Aamir Ali}
\author[e]{Yuji Chinone}
\author[f]{Kevin Crowley}
\author[g]{Giulio Fabbian}
\author[h]{Patricio A. Gallardo}
\author[d]{Neil Goeckner-Wald}
\author[i]{Brian Keating}
\author[h]{Brian Koopman}
\author[j,k]{Akito Kusaka}
\author[k]{Frederick Matsuda}
\author[m,n]{Philip Mauskopf}
\author[b]{Jeff McMahon}
\author[n,o]{Federico Nati} 
\author[p]{Giuseppe Puglisi}
\author[q]{Christian L Reichardt}
\author[r]{Maria Salatino}
\author[n]{Zhilei Xu}
\author[n]{Ningfeng Zhu}
\affil[a]{School of Electrical, Computer \& Energy Engineering, Arizona State University, Tempe, AZ, USA}
\affil[b]{Department of Physics, University of Michigan, Ann Arbor, MI, USA}
\affil[c]{Oskar Klein Centre for Cosmoparticle Physics, Stockholm University, Stockholm, Sweden}
\affil[d]{Center for Astrophysics \& Space Sciences, University of California San Diego, La Jolla, CA, USA}
\affil[e]{Department of Physics, University of California, Berkeley, Berkeley, CA, USA}
\affil[f]{Department of Physics, Princeton University, Princeton, NJ, USA}
\affil[g]{Institut d’Astrophysique Spatiale, CNRS (UMR 8617), Univ. Paris-Sud, Universit\'e Paris-Saclay, Orsay, France}
\affil[h]{Department of Physics, Cornell University, Ithaca, NY, USA}
\affil[i]{Department of Physics University of California San Diego, La Jolla CA, USA}
\affil[j]{Physics Division, Lawrence Berkeley National Laboratory, Berkeley, USA}
\affil[k]{Kavli IPMU (WPI), The University of Tokyo, Kashiwa, Chiba, Japan}
\affil[l]{School of Earth and Space Exploration, Arizona State University, Tempe, AZ, USA}
\affil[m]{Department of Physics, Arizona State University, Tempe, AZ, USA}
\affil[n]{Department of Physics \& Astronomy, University of Pennsylvania, Philidelphia, PA ,USA}
\affil[o]{Department of Physics, University of Milano-Bicocca, Milan, Italy}
\affil[p]{Department of Physics, Stanford University, Stanford, California, CA, USA}
\affil[q]{School of Physics, University of Melbourne, Melbourne, Australia}
\affil[r]{AstroParticle and Cosmology (APC) Laboratory, Paris Diderot University, Paris, France}

\authorinfo{Send correspondence to S.A.B.: sean.a.bryan@asu.edu and S.M.S.: smsimon@umich.edu}
\begin{document}
\maketitle

\begin{abstract}
The Simons Observatory (SO) is a set of cosmic microwave background instruments that will be deployed in the Atacama Desert in Chile. The key science goals include setting new constraints on cosmic inflation, measuring large scale structure with gravitational lensing, and constraining neutrino masses. Meeting these science goals with SO requires high sensitivity and improved calibration techniques. In this paper, we highlight a few of the most important instrument calibrations, including spectral response, gain stability, and polarization angle calibrations. We present their requirements for SO and experimental techniques that can be employed to reach those requirements.
\end{abstract}


\keywords{Simons Observatory, cosmic microwave background, calibration}

\section{INTRODUCTION}
\label{sec:intro}  
The Simons Observatory (SO) will observe the cosmic microwave background (CMB) in temperature and polarization to set constraints on inflation, dark matter, dark energy, and the mass and number of neutrinos. The linear polarization of the CMB can be decomposed into even (E-mode) and odd  (B-mode) parity components. If inflation occurred, it would have produced gravitational waves. Both E-modes and B-modes can be produced by gravitational waves, but E-modes are also sourced by density perturbations, so B-modes are the clearest signature of inflation~\cite{Seljak_1997, Kami_1997}. The relative size of the inflationary gravitational signal to that of the density perturbations is quantified by the scalar-to-tensor ratio $r$, which is currently constrained as $r<0.07$ at 95\% confidence \cite{bicep16}. The B-mode signal has an additional contribution from E-modes that are gravitationally lensed into B-modes by large scale structure~\cite{lensing}, which has been measured by several experiments\cite{polarbear17,act17,spt18,planck16,bicep16b}. Precision measurements of lensed B-modes are necessary to further probe neutrinos, dark energy, and dark matter. The broader CMB community has come together around these and other important science goals, as reviewed in the CMB-S4 Science Book \cite{cmb-s4-science-book}. The lensed B-mode signal peaks on small angular scales but still has a significant signal where the primordial B-mode signal would peak on degree-angular scales (multipole moments of $\ell \sim 100$). To measure the primordial B-modes, the lensed B-mode signal must be measured and removed, which requires observations on both large and small scales. Additionally, polarized foregrounds from Galactic dust and synchrotron emission can create spurious B-modes, but these foregrounds have a different spectral energy distribution than the CMB. They can thus be removed if they are well characterized across several frequencies. SO will employ a 6~m crossed-Dragone large-aperture telescope (LAT) and several 0.5~m small-aperture telescopes (SATs) to cover a wide range of angular scales and frequencies (27-270~GHz). Current progress in CMB measurements depends on developing increasingly sensitive instruments. With over 60,000 detectors, SO will be on the forefront of sensitive CMB experiments and will serve as a valuable stepping stone to CMB-S4, which will field on order 500,000 detectors.

To fully leverage this increased sensitivity, the control of calibration and systematic effects must also improve. In this paper, we describe a few of the most important calibration requirements we have identified for SO and how we plan to meet these requirements through instrument design and calibration hardware development. We highlight calibration requirements that have become newly important due to the increased sensitivity of SO. These are the requirements on the spectral response of the instrument (Sec.~\ref{sec:bandpass}), the gain stability (Sec.~\ref{sec:gain_stability}), and the absolute polarization angle (Sec.~\ref{sec:pol_angle}). This paper is part of a series of papers on the systematic and calibration studies for SO~\cite{salatino18,crowley18,gallardo18}. We are combining the detailed results of the full SO systematics and calibration studies into a comprehensive study that will be released in a series of future papers to the community for use in developing future CMB experiments such as CMB-S4.


\section{Spectral Response Calibration} \label{sec:bandpass}

\begin{figure}
\begin{center}
\begin{tabular}{c}
\includegraphics[width=0.65\textwidth]{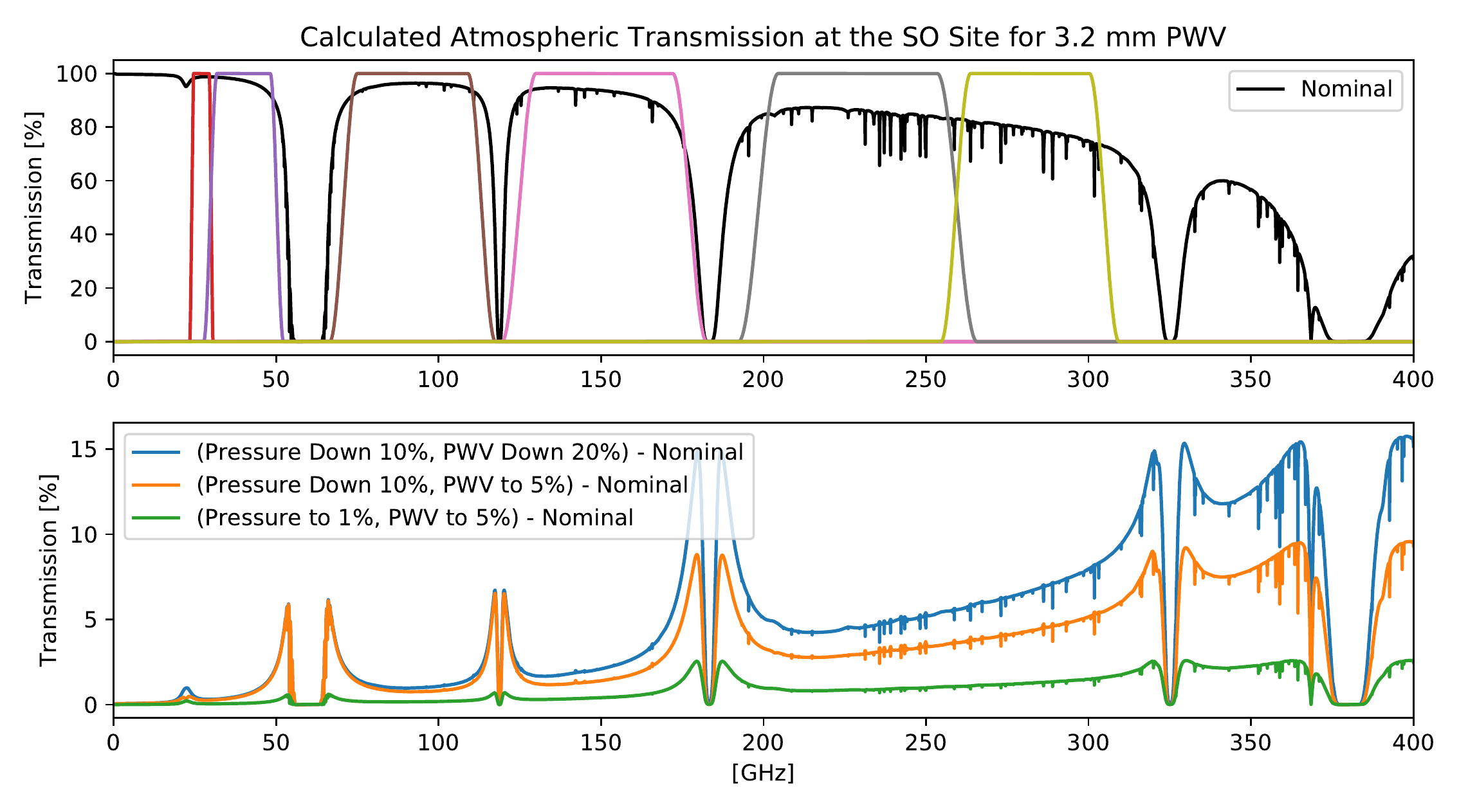}
\end{tabular}
\end{center}
\caption[example] 
{ \label{fig:spectrum} 
Calculated atmospheric transmission spectrum on the fifth-percentile worst day (3.2~mm PWV) during the best season of the year at the SO site. This corresponds to the 3~mm PWV cut implemented by several CMB experiments in the Atacama in the data selection. The bottom panel shows changes in atmospheric transmission resulting from variation in pressure and PWV conditions, and the corresponding shift in band centroids are shown in Table~\ref{tab:centroids}. If PWV or pressure measurement error is at the same level as variation shown here, that propagates into uncertainty in the band centroids.}
\end{figure}

To recover the primordial B-mode signal, the center frequency, bandwidth, and spectral shape of each frequency band must be accurately and precisely characterized. An incorrect measurement of the instrument passbands can result in incorrect foreground separation, leading to foreground residual error in the inferred B-mode signal. Understanding the end-to-end spectrum of each frequency band requires accounting both for effects inside the instrument itself and atmospheric transmission, which can vary significantly with weather conditions. Figure~\ref{fig:spectrum} shows the atmospheric transmission for a particular set of weather conditions, as well as the preliminary frequency bands for SO. As the precipitable water vapor (PWV) and atmospheric pressure change, so does the atmospheric transmission. This results in gain fluctuations and time-varying bandwidth and center frequency in each passband. Simulations with realistic passbands and atmospheric distributions for the site in Chile show that to reach the SO target for constraining $r$, the average center frequency of each band must be known to 0.5\% \cite{ward18}. This is a significantly stricter requirement than needed for past experiments.

Both improved Fourier Transform Spectrometer (FTS) measurements of the instrument and improved measurements of the atmospheric transmission spectrum are needed to meet this stricter band centroid requirement. In this section, we discuss how we will improve both the FTS and the atmospheric transmission measurements. Data from a 180 GHz weather radiometer and a temperature/pressure ground weather station will drive an atmospheric transmission model. This model will then be combined with the FTS measurements to monitor the band centroid as it changes under varying atmospheric conditions. The simulated centroid reconstruction errors are shown in Table~\ref{tab:centroids}, and meet the SO requirements.

\subsection{Measurements of the Passbands}

The spectral response of the instrument itself will be measured with an FTS. Past FTS measurements have achieved calibration accuracy to $\sim$2\%\cite{ward18}. To meet the 0.5\% center frequency requirement, we therefore expect that this uncertainty will need to improve by a factor $\geq$4. We plan to achieve this through a combination of measuring more detectors on each array, improving the coupling optics between the FTS and the optics tubes, and improving the systematics in the FTS system we are developing. Before fielding the instrument, we will make FTS measurements with a test cryostat with a full optics tube. We will also make FTS measurements in situ in the field. In both cases we will employ beam-filling coupling optics to mitigate measurement systematics from uneven illumination. For the LAT, the spectra will be measured at the entrance aperture to the camera, so the effects of the primary and secondary mirrors on the spectra will not be measured directly. However, the optics themselves are thought to have little variation with frequency since they operate by reflection from a good conductor. The impact of the varying aperture efficiency with frequency is thought to be relatively straightforward to model, and will be informed by mapping the beams at the entrance aperture to the camera as well. On the other hand, because the SATs are more compact, the FTS will notionally be able to illuminate the entire primary optic and therefore measure the end-to-end instrument passband directly. On-sky data from the SAT will be used as an independent cross check on how well the combination of FTS measurement and beam modeling worked for the LAT.


\subsection{Atmospheric Modeling}

One method of characterizing the atmosphere is through atmospheric modeling coupled with PWV and atmospheric pressure measurements. We plan to use the {\verb am }\cite{paine18} code to model the atmospheric absorption spectrum. Based on a combination of local site weather data and the NASA MERRA-2\footnote{gmao.gsfc.nasa.gov/reanalysis/MERRA-2} satellite dataset, {\verb am } has atmospheric profiles for the SO site for each of the four seasons of the year. To bound the requirements for SO, here we consider the worst-case weather conditions that could potentially still be used in the final science results. The 5\% worst weather profile during the best season of the year is 3.2~mm of PWV. Several experiments in the Atacama institute a data selection cut above 3~mm PWV for science analysis. This motivates our use of a 3.2 mm~PWV atmospheric model in this study, shown in Figure~\ref{fig:spectrum}. Idealized instrument passbands are also shown for reference.

To simulate the effect of poorly measured atmospheric conditions, we recalculated the atmospheric model with both PWV and atmospheric pressure variations. Shifting the PWV by 20\% and the atmospheric pressure by 10\%, which are large amounts for a roughly hour long observation, generates up to a 10\% change in atmospheric transmission. We then calculate the new effective band center of each band and the fractional shift due to these simulated unmeasured PWV and pressure changes. The resulting band center error is shown in Table~\ref{tab:centroids}. Even with a large 20\% PWV and 10\% pressure shift, the band centroids shifted by at most 0.4\%, which is near the required 0.5\% center frequency requirement. The weather station baselined for SO will have a radiometer sensitive to the 183~GHz water line to measure the PWV to approximately 5\%, and a barometer to measure the atmospheric pressure to approximately 1\%. Simulating atmospheric shifts at this level, the worst case uncertainty in the center frequency is 0.075\%. This is well within the requirement for SO. Since these simulations show that the combination of a weather station and 183-GHz radiometer meets the requirements for SO, and the team has experience fielding similar weather stations, this is our baseline solution for treating atmospheric passband effects.

\begin{table}
\caption{Calculated residual band centroid shifts due to atmospheric measurement error. An uncorrected 10\% pressure error and 20\% PWV error would use a significant portion of the margin on the 0.5\% band centroid requirement, but adding a PWV measurement accurate to 5\% and a pressure measurement to 1\% is achievable and exceeds the requirement with comfortable margin. In all cases, the 27 GHz band centroid uncertainty is calculated to be negligible.} 
\label{tab:centroids}
\begin{center}       
\begin{tabular}{r|lllll}
                                  & 39 GHz  & 90 GHz  & 150 GHz & 220 GHz & 270 GHz \\ \hline \hline
Pressure to 10\%, PWV to 20\% & 0.075\% & 0.001\% & 0.184\% & 0.026\% & 0.058\% \\
Pressure to 10\%, PWV to 5\%    & 0.074\% & 0.009\% & 0.112\% & 0.018\% & 0.038\% \\
Pressure to 1\%, PWV to 5\%       & 0.008\% & 0.002\% & 0.036\% & 0.005\% & 0.011\% \\ \hline \hline 
\textit{(Maximum Permissible from Requirements)}            & 0.500\% & 0.500\% & 0.500\% & 0.500\% & 0.500\% \\
\end{tabular}
\end{center}
\end{table} 

\subsection{Directly Measuring Atmospheric Transmission}

Based on the simulations shown in the previous subsection, atmospheric modeling with a single water line measurement for PWV, as well as a thermometer and barometer to measure the ground weather conditions, are baselined for SO. However, as sensitivity increases for future experiments like CMB-S4, the need to understand the instrument passbands for foreground removal will also increase, in turn increasing the need to understand the atmospheric transmission. In anticipation of these future requirements, in this section we describe how a more sophisticated multiband atmospheric sounding radiometer could be used to directly measure the atmospheric absorption spectrum without needing any atmospheric modeling.

For improved weather monitoring, we are developing a radiometer system that has many $\sim$1~GHz passbands and spans the entire single-moded bandwidth of rectangular waveguide. This will allow us to measure the atmospheric line properties at all frequencies, corresponding to different altitudes due to pressure broadening \cite{janssen93}. Some of the authors have a patent-pending\cite{bryan17} radiometer technology that could be used to cover the relevant bands for ground-based CMB instruments using WR15, WR10\cite{bryan15a}, WR6, WR5\cite{bryan16}, and WR2.8 rectangular waveguide bands. Low noise amplifiers are available commercially at all bands except WR5, but groups at the NASA Jet Propulsion Laboratory (JPL) regularly produce WR5 band amplifiers. This set of sensors would simultaneously cover every band edge likely to be used in CMB-S4, directly measuring the atmospheric attenuation without relying on any model and simultaneously improving {\verb am } models.

\begin{figure}
\begin{center}
\begin{tabular}{c}
\includegraphics[width=0.65\textwidth]{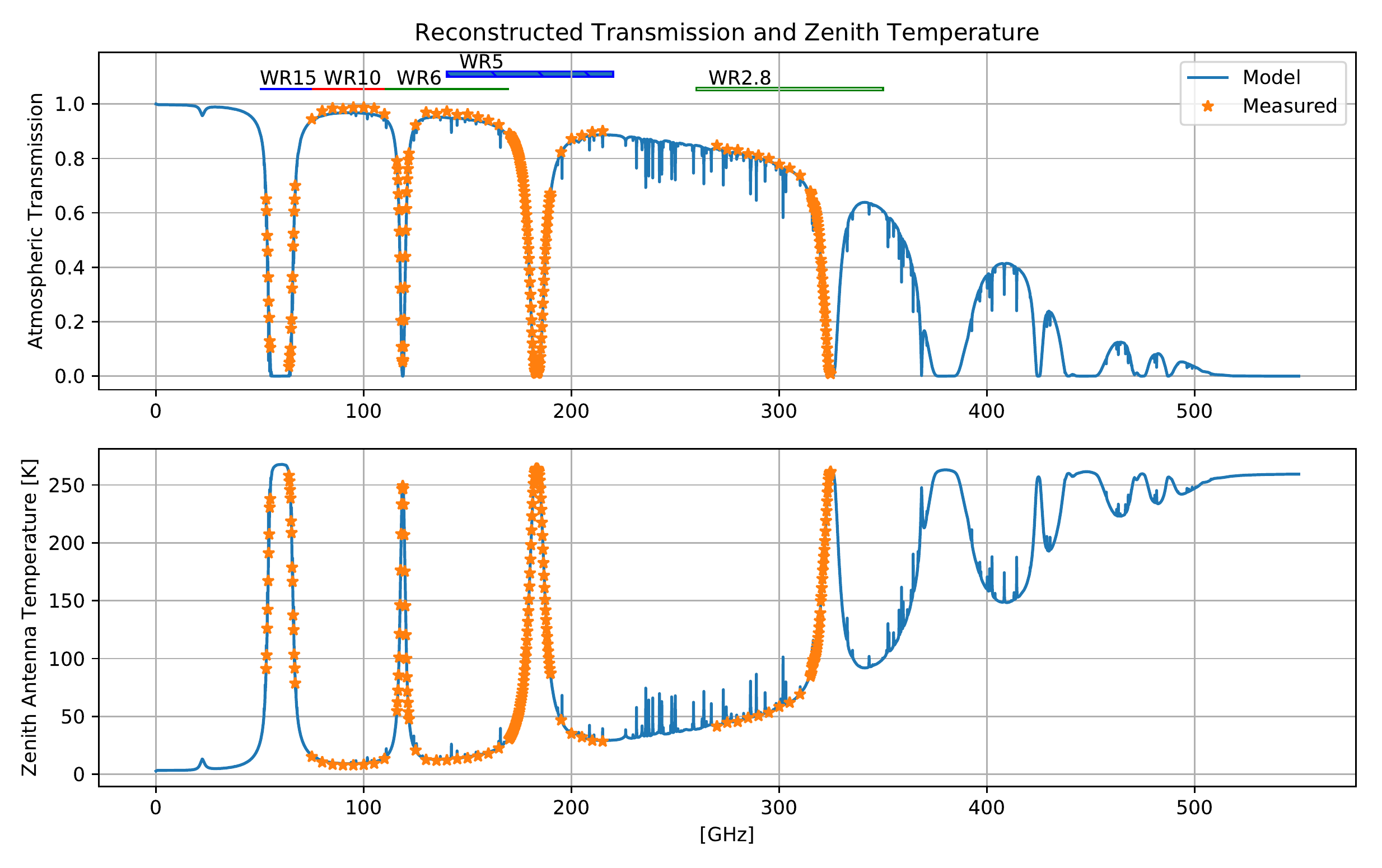}
\end{tabular}
\end{center}
\caption[example] 
{ \label{fig:direct_meas} 
An atmospheric radiometer with $\sim$1 GHz spectral resolution and wide instantaneous bandwidth (filling each of the waveguide bands highlighted in blue, red, and green above) can directly measure (orange points) the atmospheric transmission as a function of frequency without any dependence on atmospheric modeling. Our simulated measurement agrees with the input model (blue line) and, in less than a minute of data acquisition, the error on each point is 0.1\% or better depending on the optical depth at that frequency channel.}
\end{figure}

For a radiometer looking through the atmosphere at an angle $\theta$, the detected power at the radiometer frequency channel $f$ is
\begin{equation}
P_{det}(f) = g(f) \left( T_{sys}(f) + T_{atm}(f)(1-e^{-\tau(f) \sec{\theta}}) + T_{bg}(f) e^{-\tau(f) \sec{\theta}} \right).
\end{equation}
The angle $\theta$ from zenith is measured directly by the tipper mount, but all of the other quantities are unknown as a function of frequency. These are the temperature to power responsivity $g$, the internal noise temperature $T_{sys}$, the atmospheric temperature $T_{atm}$, the effective sky background temperature $T_{bg}$, and the atmospheric absorption $\tau$. By taking data looking at two beam-filling, known-temperature loads and observing the atmosphere at a range of angles, there is enough data to obtain best-fit estimates of all of these parameters without degeneracies. To roughly illustrate how the fitting code converges given this dataset, looking at the two temperature loads is a Y-factor measurement to obtain the gain $g$ and internal noise temperature $T_{sys}$. Next, looking vertically to the zenith sees nearly only the sky background $T_{bg}$, and looking horizontally sees nearly only the optically-thick horizon atmospheric temperature $T_{atm}$. Finally, the scaling with angle between horizon and zenith indicates the atmospheric absorption $\tau$. This approach yields a direct measurement of the atmospheric absorption spectrum without reference to weather data, a barometer, {\verb am } or any other modeling.

Taking expected noise levels, we simulate the best fit shown in Fig.~\ref{fig:direct_meas}. In one minute of data observing a 3.2 mm~PWV atmosphere, our simulation shows that the atmospheric transmission error in each $\sim$1~GHz bin is 0.1\% or better depending on the optical depth at that frequency bin. We take this residual noise level to generate noise realizations, perform cubic interpolation between the points, and simulate the residual band center frequency error. In the worst case scenario in the 39~GHz band, which has limited data with our proposed radiometer band selection, the band center bias with this approach is 0.11\%. This is nearly five times better than the SO requirement and has no contribution from modeling uncertainty, making this approach a good candidate for future experiments like CMB-S4.

\section{Gain Stability Requirements}\label{sec:gain_stability}
The calibration of the detector signal to the on-sky signal is a critical quantity. The absolute calibration between the detector signal and actual sky flux can be corrected by calibrating to planets and existing CMB, dust, and synchrotron temperature maps from Planck and other CMB experiments. Each individual detector has a different responsivity and thus must be weighted in the maps accordingly. This flat-fielding can be performed using intermittent calibration measurements from the atmosphere and/or an external calibration source~\cite{ABS_2018,Tajima_2012}. Experiments can also experience gain variation across observing seasons, between observations, and within observations.

The individual detector responsivities fluctuate in time due to atmospheric fluctuations, temperature drifts in the instrument, and other effects. Between each roughly hour-long observation, a combination of bias steps and current versus voltage (IV) curves can be used to calibrate the gain before each measurement. Bias steps, which have been used for both SPIDER and ACT, inject small electrical pulses into the bias line and observe the resulting detector signal response to characterize the gain. IV curves measure the saturation power and properties of the superconducting transition to measure the responsivity, but these must be used sparingly because the current on the detectors can cause array heating. For instruments with a rapidly-rotating half-wave plate (HWP), the linear dependence of the second harmonic of the HWP rotation frequency with atmospheric loading parameterized by the PWV and elevation can be used to monitor the responsivity between observations~\cite{Simon_LTD_2016}. However, the detector responsivities can also vary within individual observations. These fluctuations must be calibrated and corrected to fully understand the instrument response.

One type of these gain fluctuations are random gain fluctuations across the focal plane. If we assume that the gain is equal in a detector pair, we can model the effects of random gain drifts across the focal plane between observation periods. Assuming a linear gain drift between hourly retuning as is done in the ACT analysis, we can express the instantaneous gain $g$ at time $t$ as
\begin{equation}
g(t)=1+ \Delta g \frac{t}{t_{R}} \,\, ,
\end{equation}
where $t_{R}$ is the time between tunings, and $\Delta g$ is the gain drift between observations. We can implement this in the s4cmb\footnote{J. Peloton, https://github.com/JulienPeloton/s4cmb/} time-domain systematics pipeline and randomly draw $\Delta g$ from a Gaussian distribution across the array for each observation. These simulations show that the width of the Gaussian does not significantly impact the level of systematics from this effect. Instead, the mean gain drift is the most relevant quantity and must be known at the sub-percent level. 


Other sources of time-varying gain fluctuations include correlated detector gain variations and thermal drifts across the focal plane. Correlated detector gain variations across the focal plane from atmospheric fluctuations and studies on the thermal drifts across the array are discussed in the companion detector systematic study paper~\cite{crowley18}. To monitor the instantaneous gain, we are considering a continuously modulated calibration source for the LAT that would sit in the secondary mirror. For the SAT, we are considering using the radiometer for instantaneous PWV values so that the second harmonic of the HWP can be used as a constant responsivity monitor. Results from full studies of the thermal drifts will set thermal stability requirements on the cryogenic system. We are also designing the focal planes to have  roughly $\sim$1\% of their detectors non-optically coupled. These dark detectors will be used to measure and remove temperature drifts across the focal plane.

\section{Polarization Angle Calibration}\label{sec:pol_angle}

The absolute polarization angle is defined as the polarization orientation measured by each detector in celestial coordinates. The total polarization angle of each detector is not only determined by the antenna orientation but also has contributions from the telescope optics. Miscalibration of the absolute polarization angle can convert E-modes to B-modes. This excess B-mode signal can bias the measurement of the tensor-to-scalar ratio $r$ high and introduce uncertainty in the mass and number of neutrinos calculated from the gravitationally lensed B-mode signal. 

\subsection{Polarization Angle Calibration Requirements}

A global polarization angle miscalibration of $\Delta \psi$ results in the modified $C'_{\ell}$ coefficients~\cite{keating13}:

\begin{equation}
\begin{split}
C_{\ell}^{\prime TE} &= C_{\ell}^{TE} \cos{(2\Delta \psi)} + C_{\ell}^{TB} \sin{(2\Delta \psi)}\\
C_{\ell}^{\prime TB} &= C_{\ell}^{TB} \cos{(2\Delta \psi})-C_{\ell}^{TE} \sin{(2\Delta \psi)} \\
C_{\ell}^{\prime EE} &= C_{\ell}^{EE} \cos{^{2}(2\Delta \psi)} + C_{\ell}^{BB} \sin{^{2}(2\Delta \psi)} + C_{\ell}^{EB} \sin{(4\Delta \psi)}\\
C_{\ell}^{\prime BB} &= C_{\ell}^{BB} \cos{^{2}(2\Delta \psi)} + C_{\ell}^{EE} \sin{^{2}(2\Delta \psi)} - C_{\ell}^{EB}\sin{(4\Delta \psi)}\\
C_{\ell}^{\prime EB} &= C_{\ell}^{EB}\cos{(4\Delta \psi)}-\frac{1}{2} (C_{\ell}^{EE} - C_{\ell}^{BB}) \sin{(4\Delta \psi)} \\
\end{split}
\end{equation}
In the standard cosmological model, $C_{\ell}^{EB}$ and $C_{\ell}^{TB}$ are zero. In the presence of a miscalibration in polarization angle, the E-modes leak into B-modes and the TB and EB cross spectra are non-zero. 

\begin{figure}[h!]
\centering
\begin{subfigure}{0.5\textwidth}
  \centering
  \includegraphics[width=\linewidth]{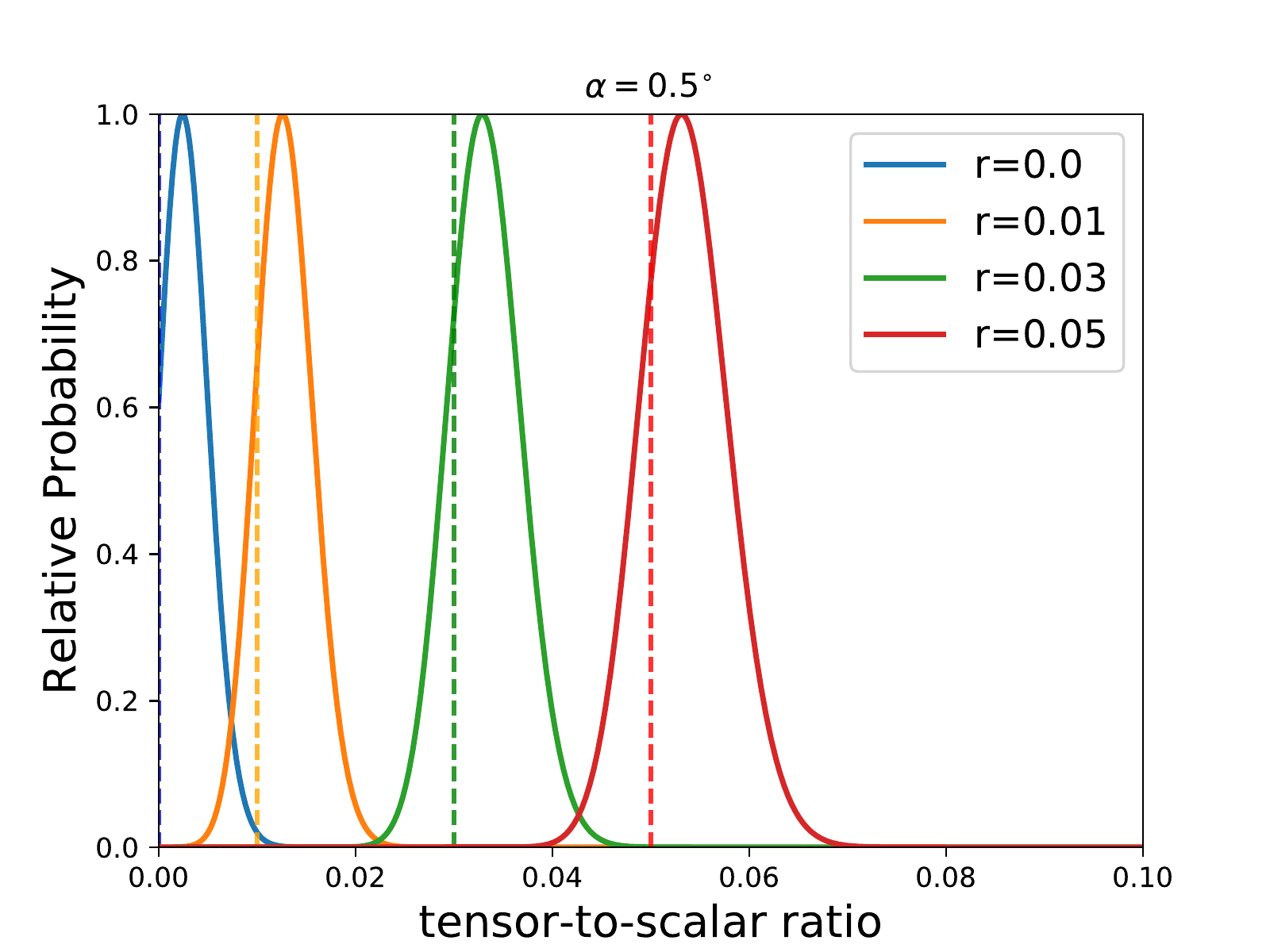}
\end{subfigure}%
\begin{subfigure}{0.5\textwidth}
  \centering
  \includegraphics[width=\linewidth]{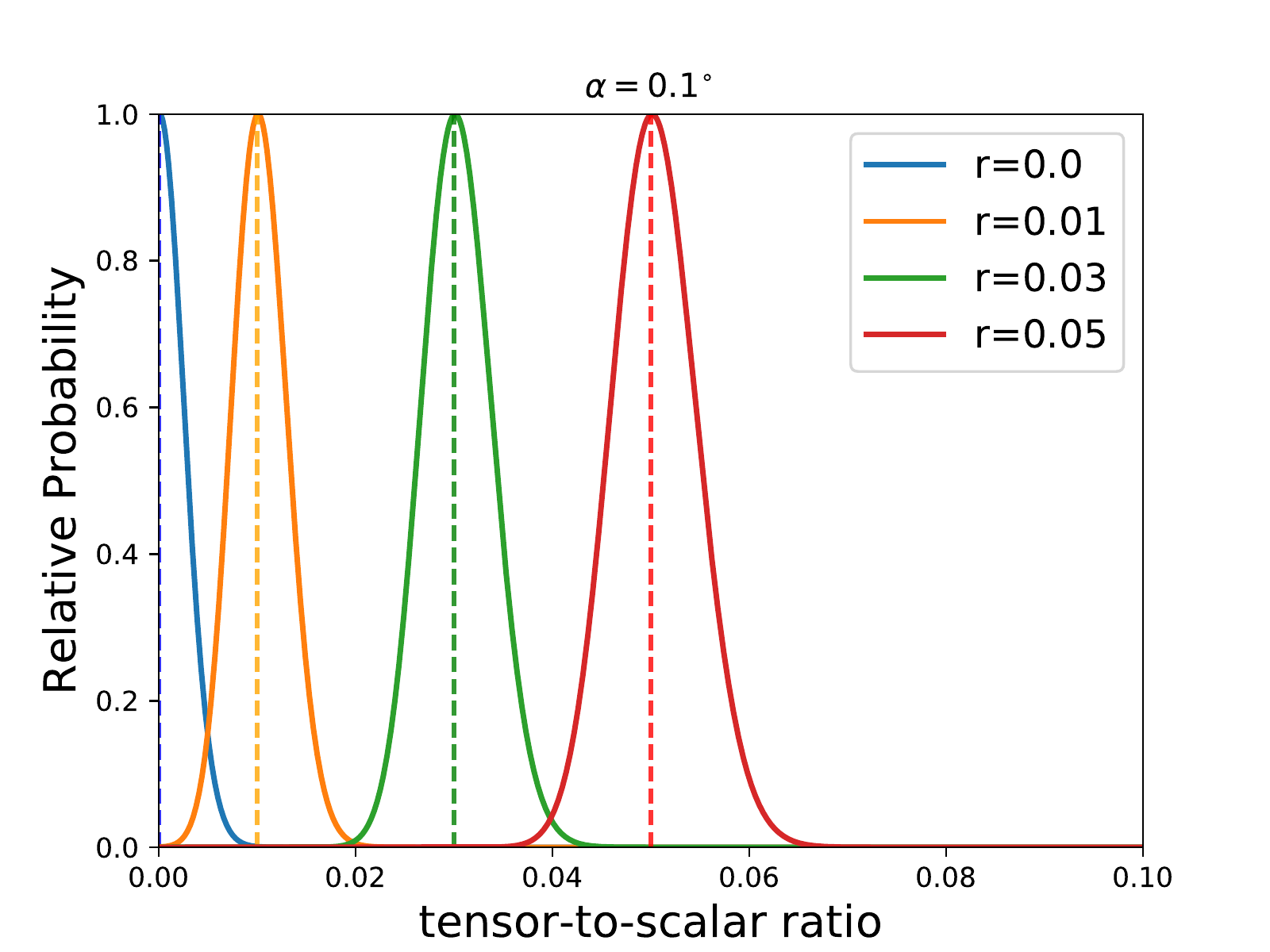}
\end{subfigure}
\caption{The probability distributions for $r$ for $r=0.05,0.03,0.01,$~and~0 in red, green, orange, and blue, respectively, for a 0.5$^{\circ}$ (left) and 0.1$^{\circ}$ (right) polarization angle uncertainty. The vertical gray lines indicate the value or $r$ where the peak of the distribution should sit. The noise is 3~$\mathrm{\mu}$Karcmin with an $\ell_{knee}=50$, which is a pessimistic case for SO. While 0.1$^{\circ}$ uncertainty introduces a negligible offset in $r$, 0.5$^{\circ}$ introduces a significant bias.}
\label{fig:pol_angle_pessimistic}
\end{figure}

\begin{figure}[h!]
\centering
\begin{subfigure}{0.5\textwidth}
  \centering
  \includegraphics[width=\linewidth]{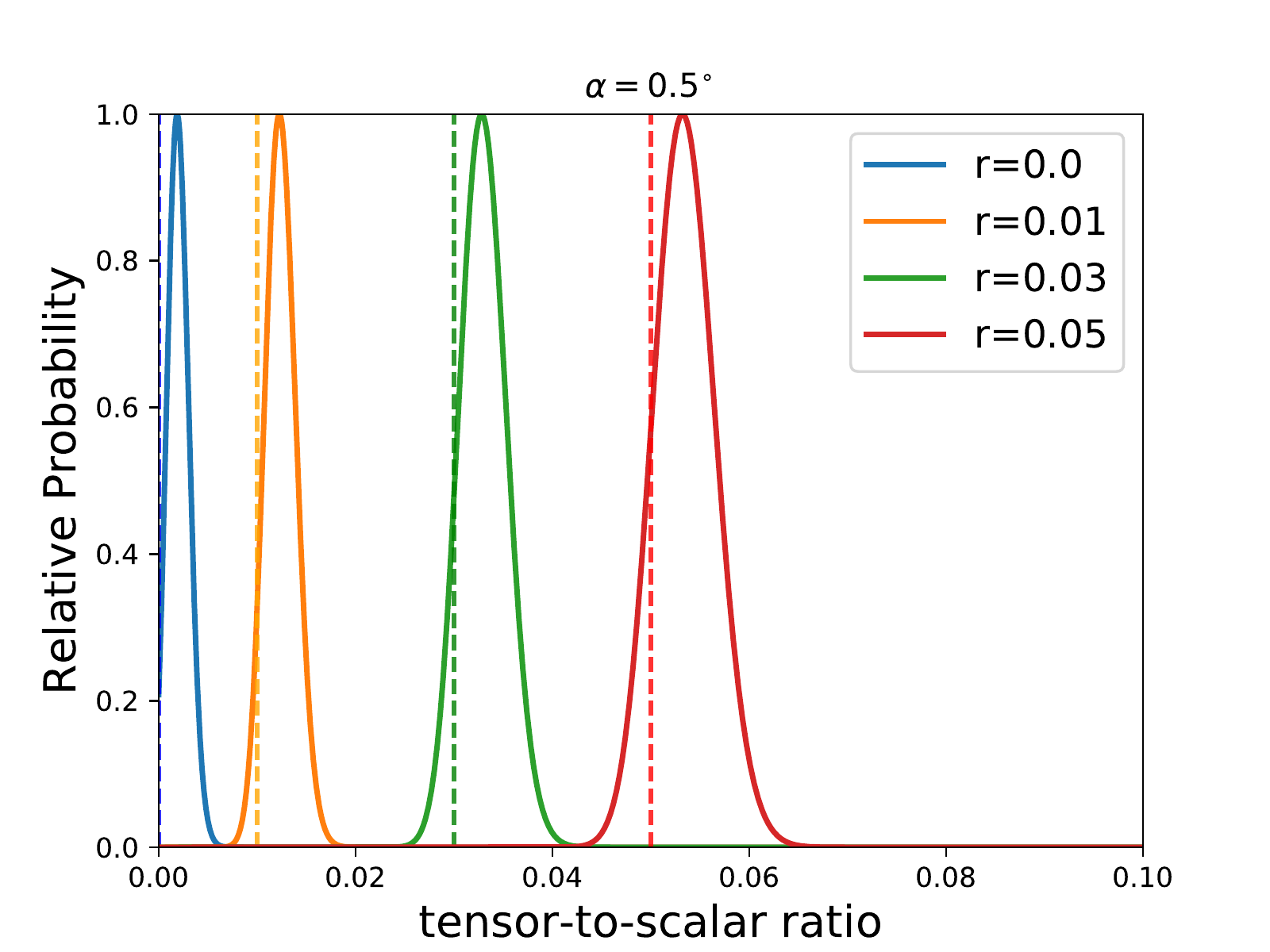}
\end{subfigure}%
\begin{subfigure}{0.5\textwidth}
  \centering
  \includegraphics[width=\linewidth]{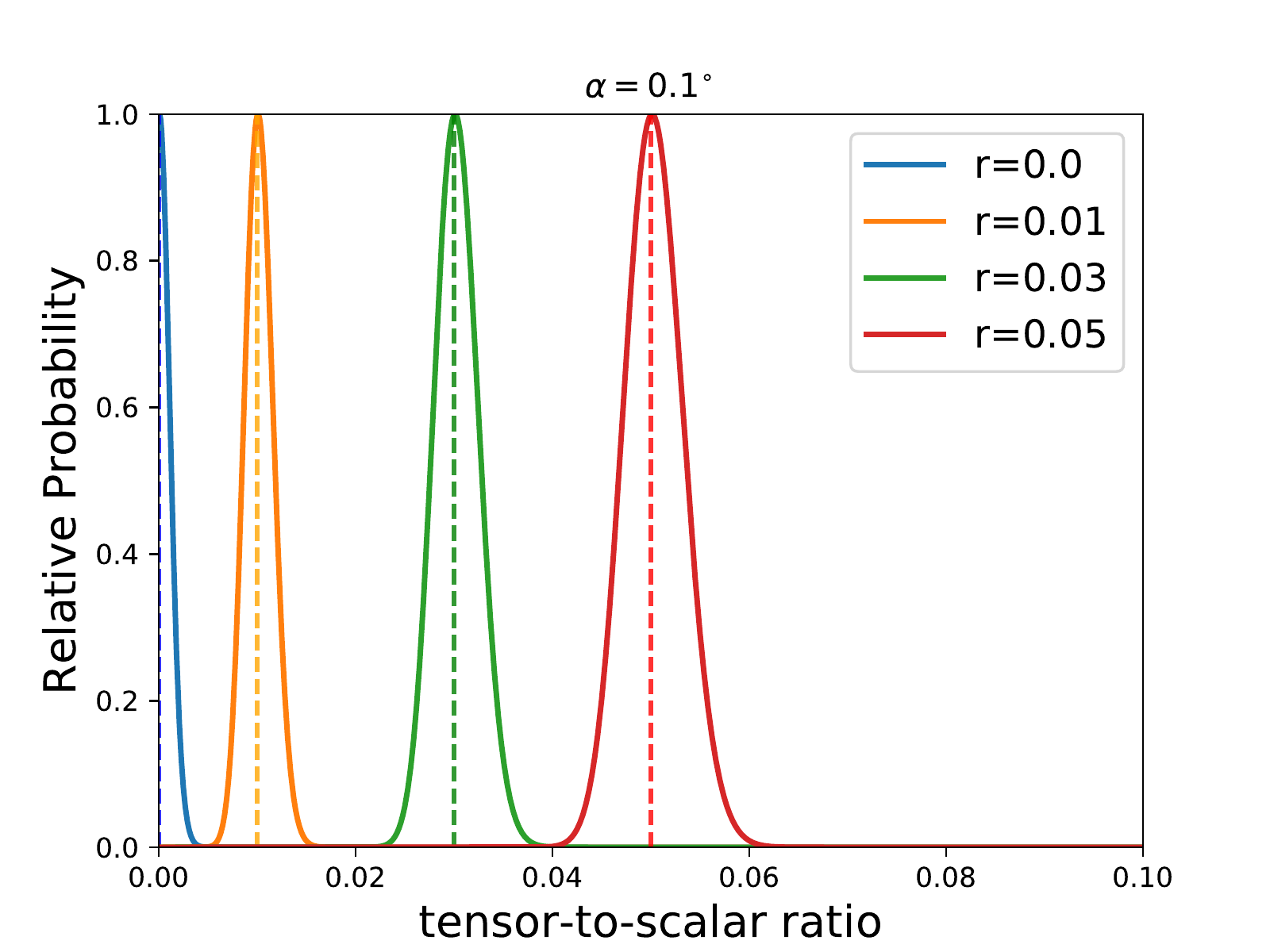}
\end{subfigure}
\caption{The probability distributions for $r$ with the same convention as Fig.~\ref{fig:pol_angle_pessimistic}.  The noise is 2~$\mathrm{\mu}$Karcmin with an $\ell_{knee}=25$, which is an optimistic case for SO. Again, 0.1$^{\circ}$ uncertainty introduces a negligible offset in $r$, 0.5$^{\circ}$ introduces a significant bias.}
\label{fig:pol_angle_optimistic}
\end{figure}

While both the gravitational lensing reconstruction (and consequently constraints on neutrino mass) from EB estimators and $r$ are affected by polarization angle miscalibration, the $r$ science goal is the driving requirement.
This requirement depends on the instrument sensitivity, delensing efficiency, and noise. Assuming a delensing efficiency of 50\%, Figs.~\ref{fig:pol_angle_pessimistic} and~\ref{fig:pol_angle_optimistic} show the probability distributions of $r$ for $r=0.05,0.03,0.01,$~and~0 in red, green, orange, and blue, respectively, from a polarization angle miscalibration for a pessimistic (3~$\mathrm{\mu}$Karcmin with an $\ell_{knee}=50$) and optimistic (2~$\mathrm{\mu}$Karcmin with an $\ell_{knee}=25$) noise forecast for SO. For all of the SO configurations, a polarization angle calibration to $\lesssim 0.1^{\circ}$ is required.

\subsection{Modeling Polarization Rotation}

There are several sources of polarization angle rotation within the optical system of a full telescope, including the detector antenna, lenses, mirrors, and HWPs. Modeling the full system together can be challenging, but simulations can give an idea of what polarization angle should be expected, which can be compared to calibration measurements.

Polarization rotation from lenses and mirrors can be modeled in CODE V\footnote{Synopsys Optical Solutions Group - https://optics.synopsys.com/codev} using a polarization-sensitive ray trace~\cite{Koopman_2016}. An input polarization is defined and propagated through the entire optical chain to the detector focal plane. The pupil averaged Stokes vector is then used to calculate the polarization angle at the focal plane. This process is repeated for 25 different fields on the sky and the results are fit to a 2D quadratic. This fit is then used to estimate the rotation for each detector on the focal plane.

In the presence of some continuously-rotating HWPs, there can also be a time-varying polarization angle rotation related to the time constant of the detectors. The finite optical time constants of the detectors can cause a phase delay with respect to the HWP rotation, which results in a polarization angle rotation~\cite{Simon_Thesis_2016, Simon_LTD2013}. If the optical time constants and the HWP frequency are known, this effect can be modeled and corrected as in the Atacama B-mode Search analysis~\cite{ABS_2018}. However, fluctuations in loading can cause time constant fluctuations both between observations and within an observation. The time constants can be measured before every $\sim$hour-long observation, but time constant fluctuations are not measured within an observation and can thus increase the uncertainty on the polarization angle.

\subsection{Available Calibration Methods}
The requirement for calibration on the polarization angle of the instrument can be met with both external calibrators and self-calibration of the instrument. External calibration methods include observations of polarized astronomical sources, wire grids, dielectric sheets, and other artificial polarized sources.

Tau~A and Cen~A have been used for absolute angle calibration in several experiments, but can only provide an accuracy between $0.5^{\circ}$ and $1^{\circ}$ due to their frequency-dependent brightness and time variability~\cite{planck16i,polarbear14,weiland11, Zemcov_2010,ABS_2018,Crab_2018}. Sparse wire grids have also been used by several experiments to measure both relative and absolute polarization angles~\cite{Tajima_2012, ABS_2018}. This technique is currently demonstrated to achieve $\sim 0.2^{\circ}$ for relative polarization angle and $\sim 1^{\circ}$ for absolute polarization angle but performance could be improved by incorporating a gravity reference into the grid. Dielectric sheets can be used similarly to the wire grid method and have reached an absolute polarization calibration uncertainty of a few degrees~\cite{ODell_2002, Moyerman_2003}. Cen~A will be too dim for calibrations on the SAT, so Tau~A will be used. While a wire grid or dielectric sheet could be used on both the LAT and SATs, in the case of the LAT, measurements would have to be performed in front of the optics tubes, which would not include the polarization angle rotations from the telescope mirrors. However, if the SATs are well-characterized and both the LAT and SAT observe the same astronomical source, the polarization angle calibration on the LAT can be improved.

Artificial polarization calibration sources with well known polarization angles can be placed in the telescope far field to measure the absolute polarization angle. The polarized sources can be mounted on a tripod, tower, drone,  weather balloon, or a CubeSat~\cite{nati_2017, Johnson2015}. The distance to the telescope far field and the telescope elevation range determines which mounting options are available. For the SO SATs, the far field ranges from $\sim 30$~m to $\sim 300$~m depending on the observation frequency, which is within the range of a tripod-, tower-, or drone-mounted polarizaton calibrator. However, the LAT far field is much farther, so the only available options for this method of calibration would be a weather balloon or a CubeSat. Ground-based sources have only achieved absolute polarization angles of 0.5$^{\circ}$, and the local gravity limits the constraining power to $\sim$20~arcsec depending on location on the mountain. However, proposed calibrators using drones and/or weather balloons where orientation is defined relative to the stars like POLOCALC could reach absolute polarization angles of $\sim$0.01$^{\circ}$~\cite{nati_2017}.

Improving external constraining power is extremely important for SO and future CMB experiments because self-calibration has several limitations. Assuming the standard model of cosmology, one can self-calibrate by using the $C_{\ell}^{\prime EB}$ and $C_{\ell}^{\prime TB}$ amplitudes and solve for the polarization angle miscalibration, which can then be corrected~\cite{keating13,kaufman14a}. However, TB and EB signals can also originate from sources beyond the standard model like cosmic birefringence and chiral gravity~\cite{Carroll_1998,Ni_1997,Carroll_1990,Choi_1999,Lue_1998}, so in the presence of additional sources of TB and EB, self-calibration is limited and can introduce biases on cosmological parameters~\cite{abitbol16,Gubitosi_2014, Pagano_2009}. Additionally, foregrounds can contaminate self-calibration and introduce biases as large as 0.5$^{\circ}$ into the absolute polarization angle calibration~\cite{Bao_2012,abitbol16}. To reach the polarization angle calibration requirement of $\sim 0.1^{\circ}$, SO will use a combination of improved external calibrators and self-calibration.


\section{Conclusions} 
SO will be one of the most sensitive CMB experiments to date. To reach the SO science goals, the calibration requirements  must be more stringent than in previous experiments, which poses a challenge to SO and future CMB experiments.

The center frequency of the detectors must be known to 0.5\%, which requires improvements in both the uncertainty of FTS measurements and the atmospheric profile. To meet this requirement, we will decrease the uncertainty in FTS measurements by a factor of 4 through characterizing systematics in the FTS and improved coupling optics. We will also improve the data driving our atmospheric transmission modeling by deploying a weather station at the SO site consisting of an atmospheric radiometer, thermometer, and barometer.

While several existing techniques can be used to calibrate the gain variation between observations, studies are underway to determine how well we will need to understand the instantaneous gain of the detectors, and we are investigating several techniques that would allow us to monitor the gain instantaneously.

Simulations show that with increased sensitivity and thus lower targets for the uncertainty in $r$, the polarization angle must be calibrated to $\sim 0.1^{\circ}$ or better. This requirement is more stringent by a factor of $\sim$5-10 than previous experiments have achieved with external calibration. While self-calibration can be used to loosen constraints on the external calibrators, it can be contaminated by foregrounds and effects not included in the standard model like cosmic birefringence. Reaching this level of polarization angle calibration requires significant development in polarization angle hardware. The calibration technologies and analyses developed for SO will represent a critical step toward the next generation of CMB experiments. 
\acknowledgments     
 
This work was supported in part by a grant from the Simons Foundation (Award \#457687, B.K.). The authors would like to thank Luca Pagano.


\bibliography{report}   
\bibliographystyle{spiebib}   

\end{document}